\documentclass[aps,prd,reprint,twocolumn,superscriptaddress,showpacs]{revtex4-2}
\usepackage{graphicx}
\usepackage{mathrsfs}
\usepackage{bm}
\usepackage{amsmath}
\usepackage{dcolumn}
\usepackage{epstopdf}
\usepackage{dsfont}
\usepackage{amssymb}
\usepackage{tabularx}
\usepackage{array}
\usepackage{float}
\usepackage{color}
\usepackage{epstopdf}
\usepackage{mathrsfs}
\usepackage[colorlinks, linkcolor=blue,anchorcolor=blue,citecolor=blue,urlcolor=blue]{hyperref}

\begin{document}
	\title{Photoinduced Odd- and Mixed-Parity Magnetism in Collinear Antiferromagnets}
	
	\author{Jin-Yang Li}
	\affiliation{School of Physics, Northwest University, Xi'an 710127, China}
	\affiliation{Institute of Modern Physics, Northwest University, Xi'an 710127, China}
	
	\author{Ruo-Yu Ning}
	\affiliation{School of Physics, Northwest University, Xi'an 710127, China}
	
	\author{Ying Zhang}
	\affiliation{School of physics and astronomy, Beijing Normal University, Beijing 100875, China}
	
	\author{Qiu-Qiu Li}
	\email{liqiuqiu@csust.edu.cn}
	\affiliation{College of Materials Science and Engineering, Changsha University of Science \textsf{\&} Technology, Changsha 410114, China}
	
	\author{Si Li}
	\email{sili@nwu.edu.cn}
	\affiliation{School of Physics, Northwest University, Xi'an 710127, China}
	\affiliation{Shaanxi Key Laboratory for Theoretical Physics Frontiers, Xi'an 710127, China}
	\affiliation{Peng Huanwu Center for Fundamental Theory, Xi'an 710127, China}
	\affiliation{Fundamental Discipline Research Center for Quantum Science and Technology of Shaanxi Province, Xi'an 710127, China}

	\author{Wen-Li Yang}
	\email{wlyang@nwu.edu.cn}
	\affiliation{Institute of Modern Physics, Northwest University, Xi'an 710127, China}
	\affiliation{Shaanxi Key Laboratory for Theoretical Physics Frontiers, Xi'an 710127, China}
	\affiliation{Peng Huanwu Center for Fundamental Theory, Xi'an 710127, China}
	\affiliation{Fundamental Discipline Research Center for Quantum Science and Technology of Shaanxi Province, Xi'an 710127, China}

\begin{abstract}
Altermagnetism, characterized by compensated collinear magnetic moments and even-parity momentum-dependent spin splitting, has emerged as a distinct form of magnetic order. However, realizing odd- and mixed-parity magnetism in fully compensated collinear magnets remains challenging. Here, we demonstrate light-induced odd- and mixed-parity magnetism in conventional collinear antiferromagnets. Combining Floquet theory with first-principles calculations, we show that circularly polarized light (CPL) induces odd-parity $f$-wave spin splitting in monolayer CoS and CoSe, whereas elliptically polarized light (EPL) induces $p$-wave spin splitting. Uniaxial strain combined with CPL offers an alternative route to the $p$-wave state. Furthermore, combining an electric field with CPL or EPL generates mixed-parity spin splitting. Light-induced breaking of $\mathcal{PT}$ symmetry, together with spin-orbit coupling, gives rise to a finite anomalous Hall response. Our findings establish a versatile approach to engineering spin splitting and controlling electronic and magnetic properties in compensated antiferromagnets.
\end{abstract}

\maketitle
\section{Introduction}
The discovery of compensated magnetic states beyond conventional ferromagnetism and antiferromagnetism has broadened the landscape of magnetic phenomena and opened new avenues for exploring unconventional magnetism~\cite{vsmejkal2022emerging,vsmejkal2022beyond,bai2024altermagnetism,song2025altermagnets,hayami2019momentum,vsmejkal2020crystal,wu2007fermi,ma2021multifunctional,yuan2020giant,smejkal2022giant,fender2025altermagnetism}. Among these emerging phases, altermagnetism represents a distinct form of collinear magnetic order that combines compensated magnetic moments in real space with momentum-dependent spin splitting in the electronic structure~\cite{vsmejkal2022emerging,vsmejkal2022beyond,bai2024altermagnetism,song2025altermagnets}. A wide variety of altermagnetic materials have been theoretically identified~\cite{ma2021multifunctional,zeng2024description,liu2024twisted,duan2025antiferroelectric,leeb2024spontaneous,gu2025ferroelectric,xu2025altermagnetic,gao2025ai,pan2024general,lin2025coulomb,chen2025electrical}, with several characteristic electronic and magnetic signatures subsequently confirmed experimentally~\cite{regmi2025altermagnetism,reimers2024direct,zhou2025manipulation,karube2022observation,bose2022tilted,fedchenko2024observation,ding2024large,han2024electrical,reichlova2024observation,lee2024broken,osumi2024observation,krempasky2024altermagnetic,mazin2023altermagnetism,hariki2024x,jiang2025metallic,zhang2025crystal,zhang2026arpes}. A defining feature of this magnetic state is its even-parity spin splitting, which can exhibit $d$-, $g$-, and $i$-wave symmetries. The rapid development of altermagnetism has further stimulated the exploration of broader classes of unconventional compensated magnets, including fully compensated ferrimagnets (fFIMs)~\cite{mazin2022altermagnetism,kawamura2024compensated,liu2025two,li2026}, odd-parity magnets with $p$- and $f$-wave spin textures~\cite{hellenes2023p,yamada2025metallic,song2025electrical}, noncoplanar antiferromagnets~\cite{takagi2023spontaneous,zhu2024observation}, and type-IV two-dimensional collinear magnets~\cite{bai2025anomalous}.

Among these unconventional compensated magnetic systems, odd-parity magnetic states are characterized by spin-split electronic structures in which the spin polarization reverses upon momentum inversion, satisfying $\varepsilon(s,\bm{k})=\varepsilon(-s,-\bm{k})$, where $\bm{k}$ denotes the crystal momentum and $s$ represents the spin index. Such odd-parity spin splitting has recently attracted increasing attention, with some recent studies exploring its possible realizations, symmetry properties, and physical consequences~\cite{hellenes2023p,yamada2025metallic,song2025electrical,brekke2024minimal,luo2025spin,chakraborty2025highly,yu2025odd,zhu2026floquet,huang2026light,li2026floquet,zeng2026odd,liu2026light,mitscherling2026microscopic,priessnitz2026ferroelectric,neumann2026odd,zhang2026odd,puzhang2026odd,bai2026time,liu2026complete,fu2026complete}. Compared with the well-established even-parity spin splitting in altermagnets, however, odd-parity spin splitting remains less explored, particularly with regard to its symmetry properties and external-field tunability. This motivates the search for versatile approaches to control the parity and angular character of spin splitting in compensated magnetic systems. A key question is whether external driving fields can systematically tune different odd-parity spin-splitting states and their associated momentum-space spin textures while preserving the underlying collinear magnetic order. Moreover, the interplay of different external perturbations may provide access to more complex spin-splitting patterns beyond the pure odd-parity states.

\begin{figure*}[htb]
	\includegraphics[width=16.5cm]{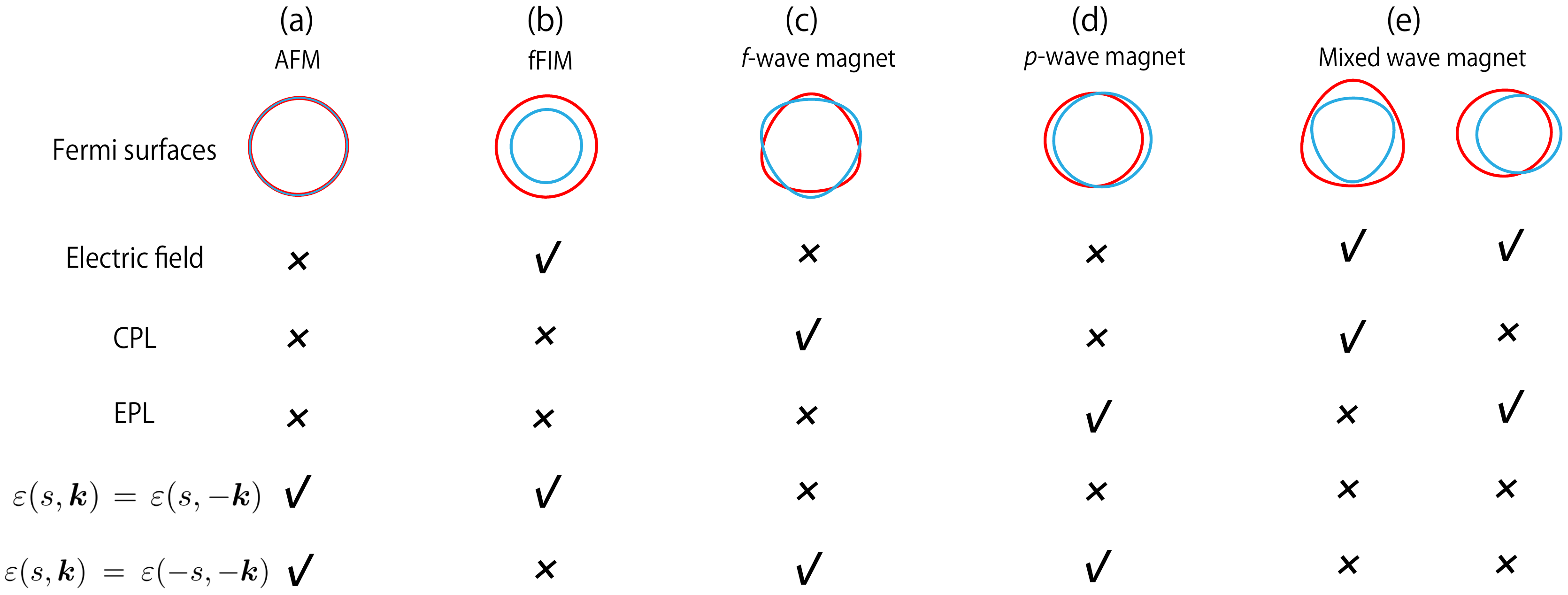}
	\caption{Various compensated magnetic states under external-field control.
		(a) Conventional collinear antiferromagnetism (AFM). (b) An fFIM state induced by applying an electric field to the AFM. (c) $f$-wave magnetism, induced by CPL irradiation of the AFM. (d) $p$-wave magnetism, induced by EPL irradiation of the AFM. (e) Mixed-wave magnetism, which can be realized by applying an electric field together with either circularly or EPL to the AFM.}
	\label{fig1}
\end{figure*}

In this work, we demonstrate that light can induce odd- and mixed-parity magnetism in conventional collinear antiferromagnets. Taking monolayer CoS and CoSe as representative examples, we systematically investigate light-induced magnetic states by combining first-principles calculations with Floquet theory. Both monolayers are collinear antiferromagnetic semiconductors with spin-degenerate band structures protected by $\mathcal{PT}$ symmetry. We find that irradiation with circularly polarized light (CPL) induces odd-parity $f$-wave spin splitting, whereas elliptically polarized light (EPL) gives rise to odd-parity $p$-wave spin splitting. Furthermore, uniaxial strain combined with CPL can transform the spin splitting into an odd-parity $p$-wave form. Remarkably, when an electric field is applied together with CPL or EPL, the spin splitting can evolve into mixed-parity forms. The external-field control scheme is illustrated in Fig.~\ref{fig1}. Because $\mathcal{PT}$ symmetry is broken under CPL or EPL irradiation, an anomalous Hall effect can emerge. Our work establishes a versatile route to realizing odd- and mixed-parity spin splitting in compensated magnetic systems and opens a new avenue for light-controlled magnetic phenomena and spintronic applications.

\section{First-principles Methods}
First-principles calculations were performed using the Vienna \textit{ab initio} Simulation Package (VASP)~\cite{kresse1994ab,kresse1996efficient} within the framework of density functional theory (DFT). The projector augmented-wave (PAW) approach was adopted to describe the electron–ion interactions~\cite{blochl1994projector}, while the exchange–correlation effects were accounted for by the Perdew–Burke–Ernzerhof (PBE) functional within the generalized gradient approximation (GGA)~\cite{perdew1996generalized}. A plane-wave cutoff energy of 450 eV was used throughout the calculations. The atomic positions and lattice parameters were relaxed until the residual forces on each atom were below 0.01 eV/\AA\ and the total-energy change was less than $10^{-6}$ eV. Brillouin-zone integrations were carried out on a $\Gamma$-centered $16\times16\times1$ $k$-point grid. A vacuum layer of 20 \AA\ was included along the out-of-plane direction to suppress interactions between periodic images. The on-site Coulomb correlations of the Co $3d$ states were incorporated within the DFT+$U$ framework~\cite{Anisimov1991,dudarev1998}, using an effective Hubbard parameter of $U=2$ eV.

To describe the coupling between the Floquet laser field and our material system, we first constructed a maximally localized Wannier-function (MLWF) representation from the plane-wave-based first-principles calculations using Wannier90~\cite{marzari1997maximally,souza2001maximally,pizzi2020wannier90}. The orbital-resolved band structures show that the low-energy states near the valence-band maximum are predominantly contributed by the Co $d$ and S/Se $p$ orbitals. Utilizing these MLWFs, we construct the real-space tight-binding Hamiltonian, expressed as
\begin{equation}   
	H = \sum_{mn} \sum_{j} t_{j}^{mn} C_{m}^{\dagger}(\mathbf{R}_{j}) C_{n}(\mathbf{R}_{0}) + \mathrm{h.c.},
	\label{eq:TB_real_space} 
\end{equation} 
where $t_j^{mn}$ denotes the hopping matrix element from the $n$-th localized Wannier orbital at site $\mathbf{R}_{0}$ to the $m$-th orbital at site $\mathbf{R}_{j}$. The operators $C_{m}^{\dagger}(\mathbf{R}_{j})$ and $C_{n}(\mathbf{R}_{0})$ correspond to the creation and annihilation operators of an electron at the respective Wannier centers. 

When driven by a time-periodic CPL characterized by the vector potential $\mathbf{A}(t) = \mathbf{A}(t+T)$, the optical coupling is introduced via the Peierls substitution into the hopping amplitudes:
\begin{equation}   
	t_j^{mn}(t) = t_j^{mn} \exp\left[ i\frac{e}{\hbar}\mathbf{A}(t)\cdot\mathbf{d}_{mn} \right],
	\label{eq:Peierls} 
\end{equation} 
where $\mathbf{d}_{mn}$ represents the spatial displacement vector linking the two Wannier orbitals. Consequently, the time-dependent Hamiltonian takes the form:
\begin{equation}   
	H(\mathbf{k},t) = \sum_{mn} \sum_{j} t_{j}^{mn}(t) e^{i\mathbf{k}\cdot\mathbf{R}_{j}} C_{m}^{\dagger}(\mathbf{k},t) C_{n}(\mathbf{k},t) + \mathrm{h.c.}.
	\label{eq:H_kt} 
\end{equation} 
Owing to the translational symmetry of the lattice and the periodicity of the driving field, Floquet theory allows us to expand the time-dependent creation and annihilation operators into the Floquet-Bloch basis:
\begin{equation}
	\begin{split}   
	C_m^\dagger(\mathbf{k},t) &= \sum_j \sum_{\alpha=-\infty}^\infty C_{\alpha m}^\dagger(\mathbf{R}_j) e^{i\mathbf{k}\cdot\mathbf{R}_j - i\alpha\omega t},\\   
	C_m(\mathbf{k},t) &= \sum_j \sum_{\alpha=-\infty}^\infty C_{\alpha m}(\mathbf{R}_j) e^{-i\mathbf{k}\cdot\mathbf{R}_j + i\alpha\omega t}.
	\end{split} 
\end{equation} 
Accordingly, the driven system can be recast into a static effective Hamiltonian defined in the extended momentum-frequency space:
\begin{equation}   
	H(\mathbf{k},\omega) = \sum_{m,n} \sum_{\alpha,\beta} \left[ H_{\alpha-\beta}^{mn} + H_{\Omega} \right] C_{\alpha m}^{\dagger}(\mathbf{k}) C_{\beta n}(\mathbf{k}) + \mathrm{h.c.}, 
\end{equation} 
with the matrix elements given by
\begin{equation}   
	H_{\alpha-\beta}^{mn} = \sum_{j} e^{i\mathbf{k}\cdot\mathbf{R}_{j}} \frac{1}{T} \int_{0}^{T} t_{j}^{mn} \exp\left[ i\frac{e}{\hbar}\mathbf{A}(t)\cdot\mathbf{d}_{mn} \right] e^{i(\alpha-\beta)\omega t} dt, 
\end{equation} 
and the diagonal photon dressing term
\begin{equation}   
	H_\Omega = \alpha\hbar\omega \delta_{mn} \delta_{\alpha\beta}, 
\end{equation} 
where $\hbar\omega$ is the photon energy, and $(\alpha, \beta)$ label the Floquet sideband indices spanning from $-\infty$ to $+\infty$. In the high-frequency regime, the driven system is conveniently captured by an effective time-independent Floquet Hamiltonian~\cite{maricq1982application,grozdanov1988quantum,rahav2003effective,rahav2003time,goldman2014periodically}:
\begin{equation}   
	H_{\mathrm{eff}}(\mathbf{k},\omega) = H_{0}(\mathbf{k},\omega) + \frac{[H_{1}(\mathbf{k},\omega), H_{-1}(\mathbf{k},\omega)]}{\hbar\omega} + \mathcal{O}\left(\frac{1}{\omega^{2}}\right). 
\end{equation}

\section{CPL-Induced $f$-Wave Odd-Parity Spin Splitting}
\begin{figure*}[htb]
	\includegraphics[width=15cm]{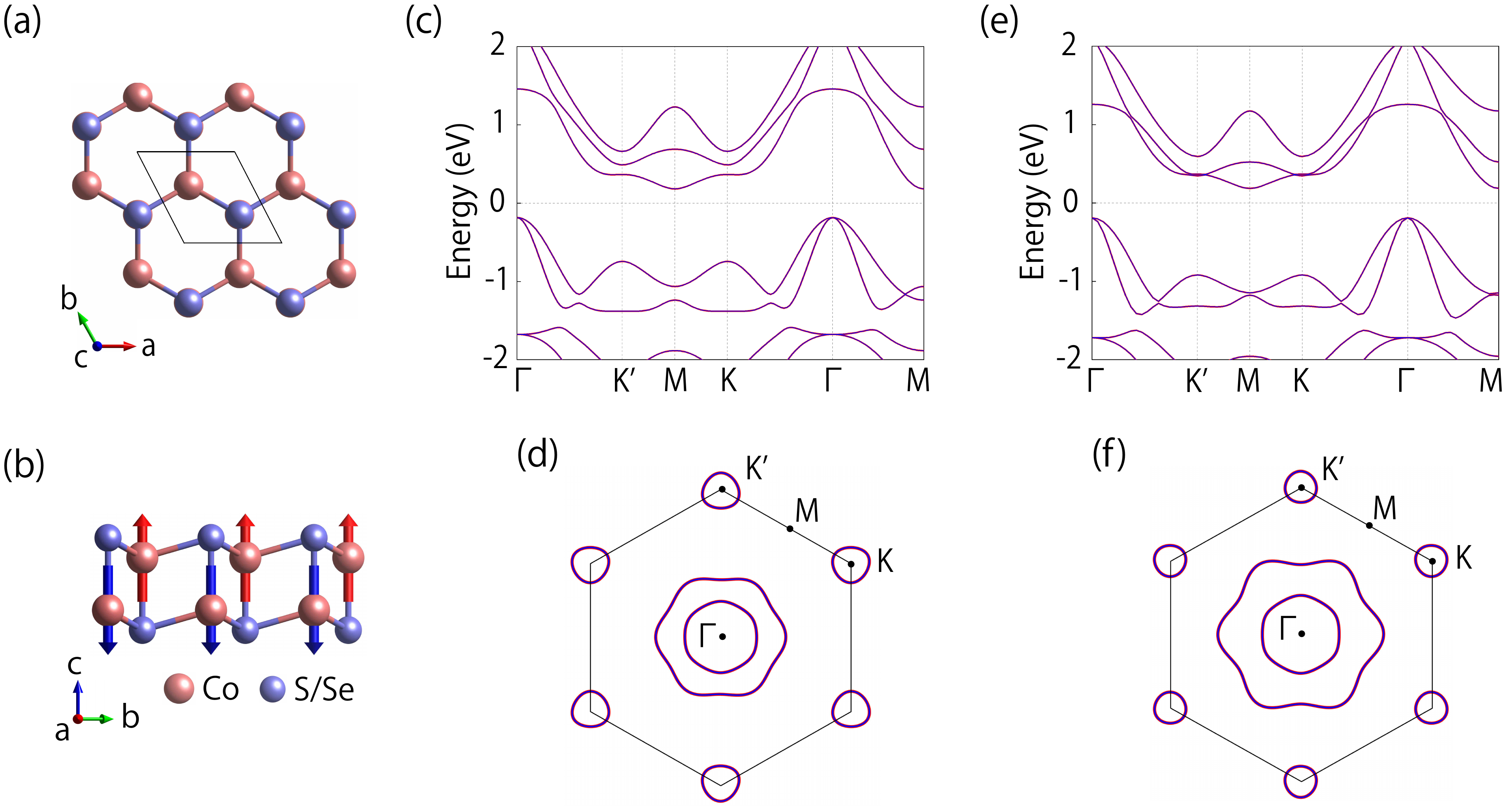}
	\caption{(a) Top and (b) side views of the crystal structures of monolayer CoS and CoSe.
		(c) Electronic band structure and (d) spin-resolved isoenergy contour of monolayer CoS at $E=-0.8$~eV.
		(e) Electronic band structure and (f) corresponding spin-resolved isoenergy contour of monolayer CoSe at $E=-0.95$~eV.
		Red and blue denote the two opposite spin channels, respectively. }
	\label{fig2}
\end{figure*}
Monolayer CoS and CoSe consist of two buckled honeycomb Co$X$
($X=\mathrm{S},\mathrm{Se}$) sublayers connected by Co--$X$ bonds,
as illustrated in Figs.~\ref{fig2}(a) and~\ref{fig2}(b). After structural
optimization, the lattice constants are $a=b=3.85$~\AA{} for CoS
and $a=b=4.03$~\AA{} for CoSe. The structural stability of these
monolayers has been established previously~\cite{li2026}. Both
systems crystallize in the $P\overline{3}m1$ space group (No.~164).
Their magnetic ground state is a N\'{e}el-type collinear
antiferromagnetic configuration, in which the magnetic moments of
the two sublattices are oppositely aligned~\cite{li2026}. The system preserves the threefold rotational symmetry $C_{3z}$ and the combined
$\mathcal{PT}$ symmetry, where $\mathcal{P}$ and $\mathcal{T}$
denote spatial inversion and time reversal, respectively. Both materials are semiconductors, and their electronic bands are spin degenerate owing to the $\mathcal{PT}$ symmetry, as shown in Figs.~\ref{fig2}(c)--(f). 

\begin{figure*}[htb]
	\includegraphics[width=17cm]{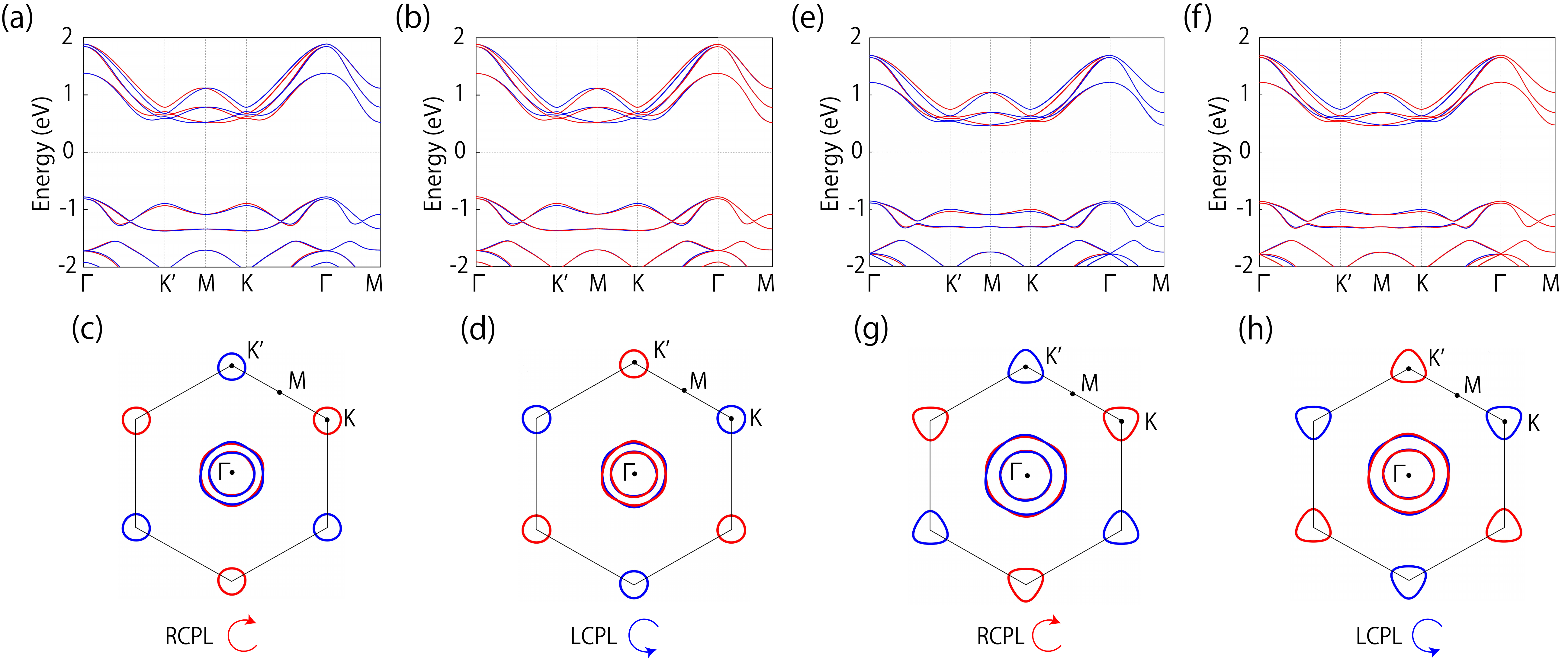}
	\caption{Electronic band structures of CoS under (a) right- and (b) left-handed CPL, together with the corresponding spin-resolved isoenergy contours at $E=-0.93$~eV shown in (c) and (d), respectively.	Electronic band structures of CoSe under (e) right- and (f) left-handed CPL, with the corresponding spin-resolved isoenergy contours at $E=-1.03$~eV shown in (g) and (h), respectively. Red and blue denote the two opposite spin channels, respectively. The photon energy and reduced vector-potential amplitude are $\hbar\omega=10$~eV and $\tilde{A}=0.3$~\AA$^{-1}$, respectively.}
	\label{fig3}
\end{figure*}
In the absence of spin–orbit coupling (SOC), the magnetic symmetries of the system are naturally described by spin-space groups, in which a general spin-space operation is written as $[R_S || R_L]$, with $R_S$ and $R_L$ acting in spin and real spaces, respectively. For a collinear AFM system, two symmetries are particularly important for relating electronic states at opposite crystal momenta. The first is $[C_{2\perp} ||\mathcal{P}]$, where $C_{2\perp}$ denotes a twofold rotation in spin space about an axis perpendicular to the spin-quantization direction, while spatial inversion $\mathcal{P}$ exchanges the two magnetic sublattices. This symmetry imposes the constraint $\varepsilon(s,\bm{k})=\varepsilon(-s,-\bm{k})$, corresponding to an odd-parity spin splitting. The second symmetry is $[C_{2\perp}\mathcal{T}||E]$, which combines time-reversal symmetry $\mathcal{T}$ with a twofold spin-space rotation $C_{2\perp}$, with $E$ denoting the identity operation in real space. It constrains the energy dispersion to satisfy $\varepsilon(s,\bm{k})=\varepsilon(s,-\bm{k})$,
giving rise to an even-parity spin splitting.
The coexistence of these two spin-space symmetries is equivalent to the conventional $\mathcal{PT}$ constraint and guarantees spin degeneracy in AFMs. Importantly, breaking $[C_{2\perp}\mathcal{T}|| E]$ while retaining $[C_{2\perp} ||\mathcal{P}]$ lifts the spin degeneracy, whereas the latter continues to enforce $\varepsilon(s,\bm{k})=\varepsilon(-s,-\bm{k})$. The resulting spin-split electronic structure thus exhibits an odd-parity character, with the spin-resolved Fermi surfaces related by simultaneous reversal of spin and crystal momentum.
Light irradiation, including CPL and EPL, provides a natural means of realizing such selective symmetry breaking. In the SOC-free limit, CPL and EPL can break the spin-preserving pseudo-time-reversal symmetry $[C_{2\perp}\mathcal{T}|| E]$ while retaining the spin-flipping inversion symmetry $[C_{2\perp} ||\mathcal{P}]$. The resulting symmetry-selective Floquet engineering lifts the spin degeneracy while preserving the opposite-spin/opposite-momentum constraint, thereby providing a route to realizing light-induced odd-parity magnetism in conventional AFMs.

We next investigate how light irradiation modifies the electronic band structures of the materials. We focus on CPL and EPL, since linearly polarized light does not induce spin splitting in conventional antiferromagnetic materials, as discussed in the
Supplemental Material (SM)~\cite{SM}. We first examine the effect of CPL. The vector potential of CPL propagating along the $z$ axis can be written as $\mathbf{A}(t)=A_0\left[\cos(\omega t), \eta\sin(\omega t), 0\right]$, where $\omega$ is the light frequency, $A_0$ is the vector-potential amplitude, and $\eta=\pm1$ denotes the helicity. For convenience, we define $\tilde{A}=eA_0/\hbar$, which has the dimension of a wavevector and will be used hereafter to characterize the amplitude of the applied light field. The periodic
drive modifies the electronic Hamiltonian through the Floquet
coupling. Under CPL irradiation, the chirality of the driving field breaks time-reversal symmetry, thereby lifting the $\mathcal{PT}$-protected spin degeneracy. This effect is demonstrated in the band structures and spin-resolved Fermi surfaces of monolayer CoS and CoSe, as shown in Fig.~\ref{fig3}. In both systems, the CPL-induced Floquet bands exhibit pronounced spin splitting, in sharp contrast to the spin-degenerate bands in the absence of light. Remarkably, the corresponding spin-resolved Fermi surfaces display a characteristic $f$-wave odd-parity pattern, arising from the underlying $[E ||C_{3z}]$ symmetry of the systems. Furthermore, reversing the helicity of the driving field from right-handed CPL (RCPL) to left-handed CPL (LCPL) reverses the direction of the spin splitting, revealing a direct correspondence between the light helicity and the sign of the induced odd-parity spin polarization.

\begin{figure*}[htb]
	\includegraphics[width=17cm]{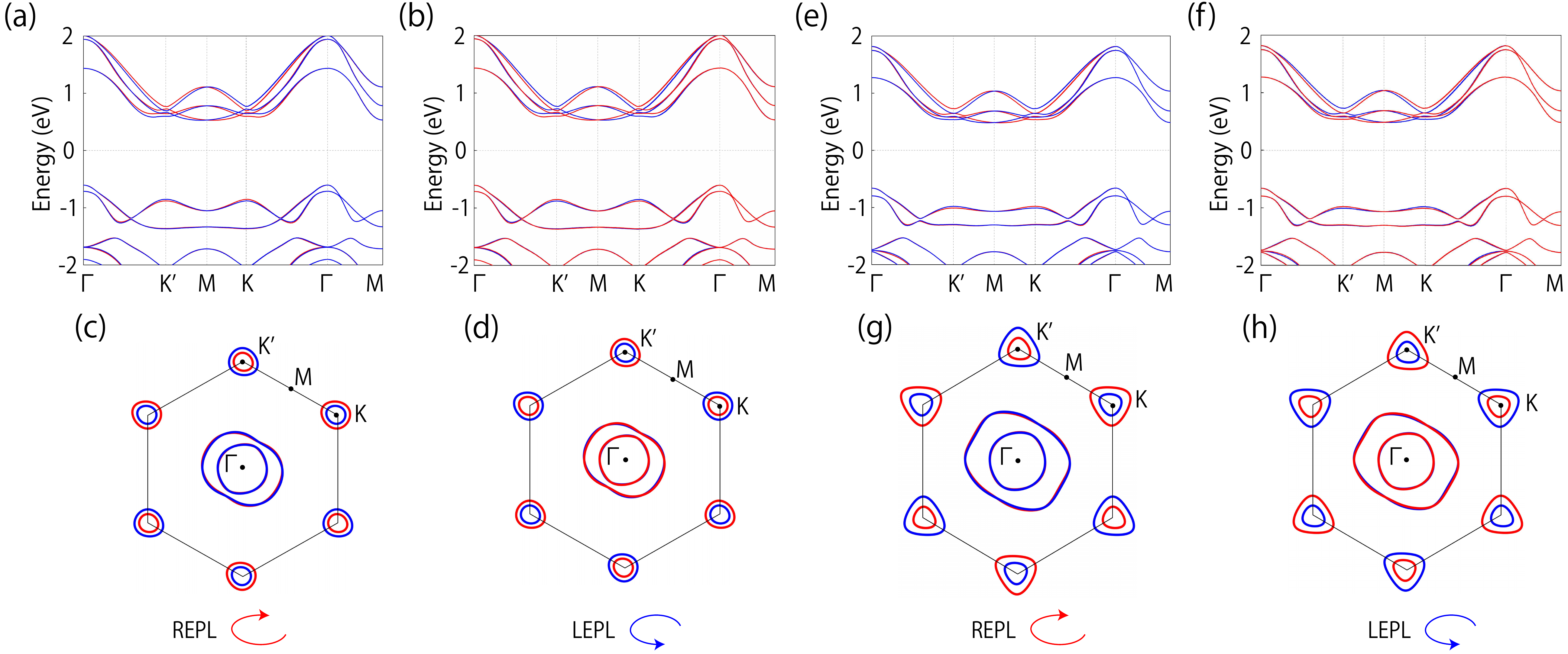}
	\caption{Electronic band structures of CoS under (a) right- and (b) left-handed EPL, with spin-resolved isoenergy surfaces at $-0.9$ eV shown in (c) and (d), respectively; and of CoSe under (e) right- and (f) left-handed EPL, with corresponding surfaces at $-1.02$ eV shown in (g) and (h), respectively. Red and blue denote spin-up and spin-down states, respectively. The calculations use $\hbar\omega=10$ eV and a light-field amplitude of $\tilde{A_{x}}=0.3$ \AA$^{-1}$ and $\tilde{A_{y}}=0.2$ \AA$^{-1}$.}
	\label{fig4}
\end{figure*}

\section{Transition from $f$-Wave to $p$-Wave}
\begin{figure*}[htb]
	\includegraphics[width=17cm]{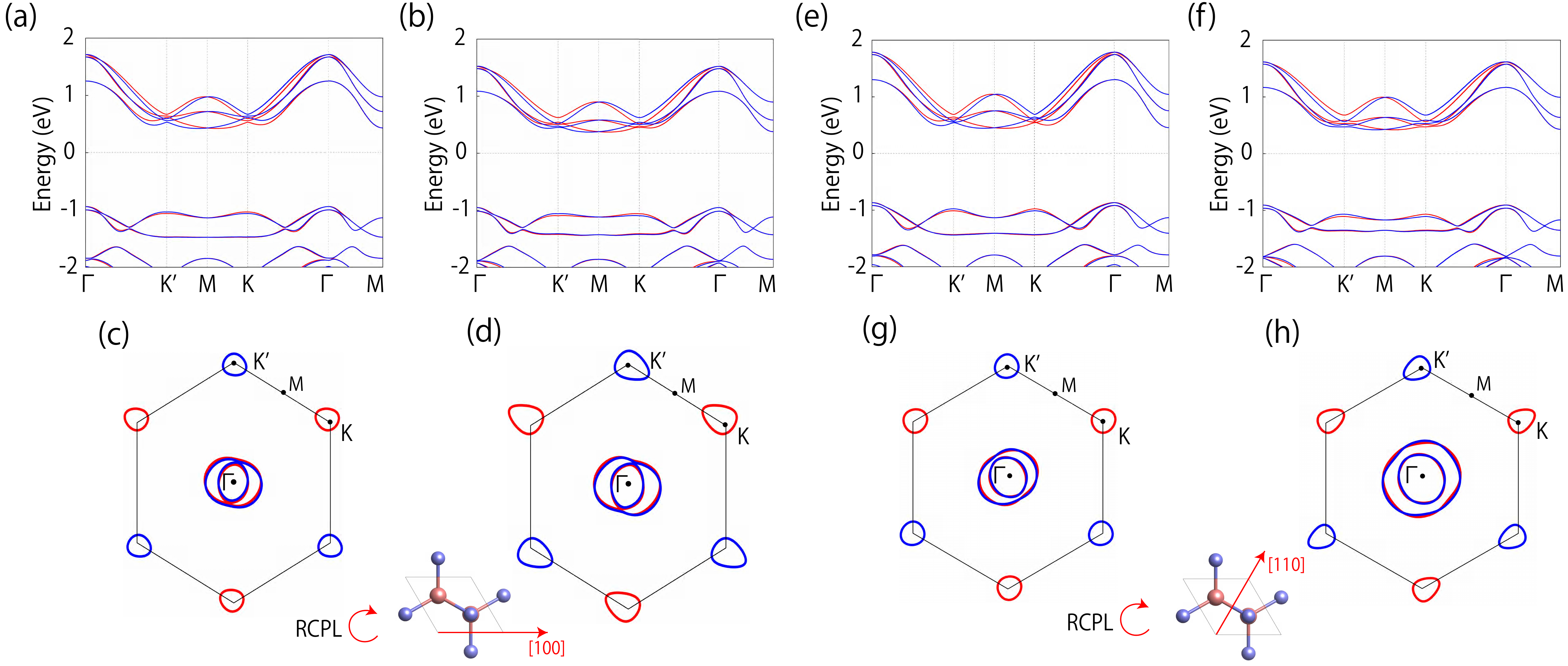}
	\caption{Electronic band structures of (a,b) CoS and CoSe under right-handed CPL and 2\% uniaxial strain along [100], with (c,d) the corresponding central-symmetric $p$-wave spin-resolved isoenergy surfaces. (e,f) Same as (a,b), but with strain along [110], and (g,h) the corresponding axial-symmetric $p$-wave spin-resolved isoenergy surfaces. Red and blue denote spin-up and spin-down states, respectively. The photon energy and vector-potential amplitude are $\hbar\omega=10$ eV and $\tilde{A}=0.3$ {\AA}$^{-1}$, respectively. The isoenergy surfaces are plotted at $-1.06$, $-1.08$, $-0.98$, and $-1.08$ eV for (c), (d), (g), and (h), respectively. }
	\label{fig5}
\end{figure*}
As discussed above, the $[E || C_{3z}]$ symmetry imposes a strong constraint on the momentum dependence of the light-induced spin splitting, leading to a characteristic $f$-wave odd-parity form. This symmetry constraint can be relaxed in two complementary ways: by modifying the polarization state of the driving field or by applying an external structural perturbation. In particular, EPL breaks the threefold rotational symmetry of the circularly driven system and thereby allows a lower-order odd-parity component to emerge. As a result, the $f$-wave spin splitting induced by CPL can be transformed into a distinct $p$-wave form under EPL. The vector potential of EPL propagating along the $z$ axis can be written as $\mathbf{A}_{\mathrm{EPL}}(t) = [A_x \cos (\omega t), \eta A_y \sin (\omega t), 0]$,
where the ratio $A_y/A_x$ characterizes the degree of ellipticity. In contrast to CPL, for which $A_x=A_y$, an anisotropic EPL field distinguishes the $x$ and $y$ directions and therefore removes the threefold rotational constraint. As shown in Fig.~\ref{fig4}, for $A_y/A_x=\frac{2}{3}$, the Floquet band structure develops a pronounced spin splitting with a characteristic $p$-wave odd-parity momentum dependence. This result demonstrates that the momentum-space structure of the light-induced spin splitting is not solely determined by the magnetic order, but can be directly controlled by the polarization state of the driving field.

The same symmetry principle can also be exploited through mechanical strain. Uniaxial strain lowers the in-plane rotational symmetry and breaks the $[E||C_{3z}]$ operation, providing an alternative route to modifying the momentum-space structure of the light-induced spin splitting. Consequently, when CPL is applied to a uniaxially strained AFM, the $[E||C_{3z}]$-protected $f$-wave form is no longer symmetry enforced, and a $p$-wave odd-parity component can emerge. To demonstrate this effect, we calculate the Floquet band structures under CPL irradiation in the presence of $2\%$ uniaxial strain along both the [100] and [110] directions, as shown in Fig.~\ref{fig5}. In both cases, the original $f$-wave spin splitting is replaced by a characteristic $p$-wave odd-parity structure. The momentum-space distribution of the spin splitting further reflects the direction of the applied strain, indicating that uniaxial strain provides an additional means of controlling the orientation and anisotropy of the light-induced spin texture.

\section{Light- and Electric-Field-Induced Mixed-Parity Spin Splitting}
\begin{figure*}[htb]
	\includegraphics[width=17cm]{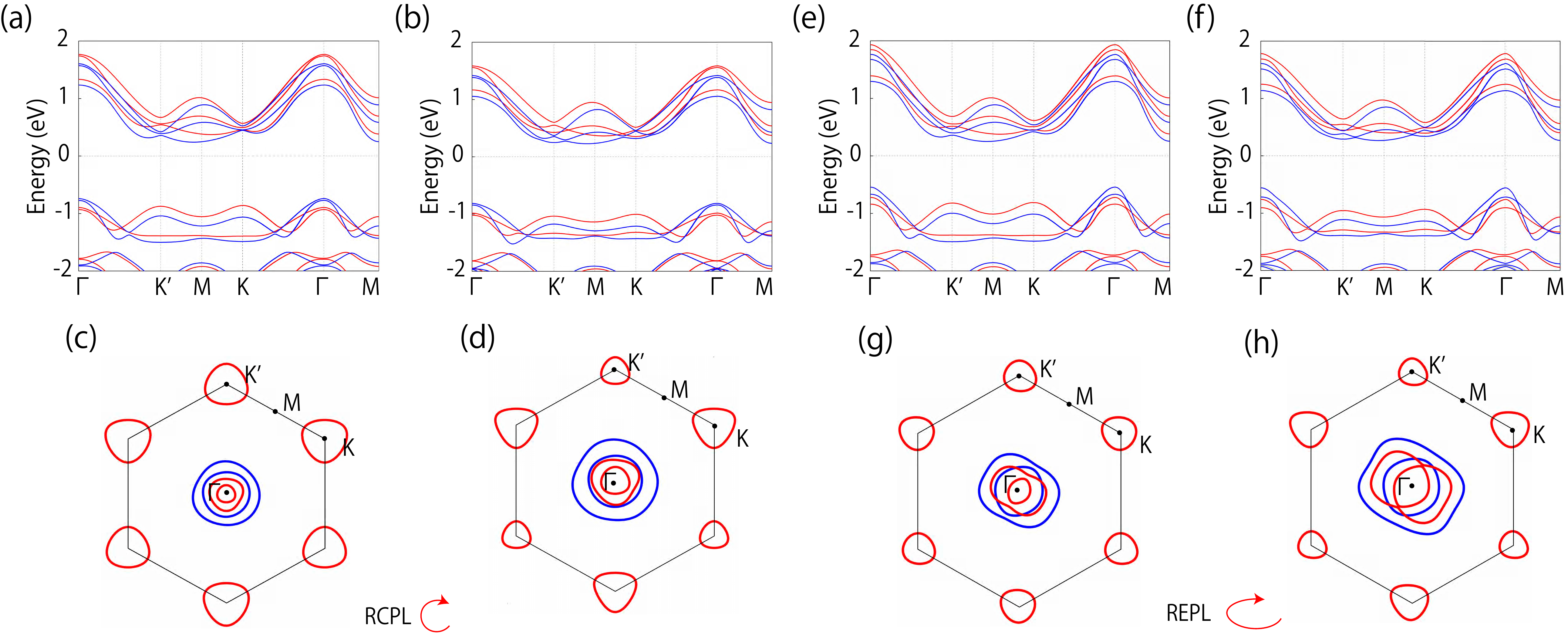}
	\caption{Electronic band structures of (a,b) CoS and CoSe under right-handed CPL and an electric field, with $\hbar\omega=10$ eV, $\tilde{A}=0.3$ \AA$^{-1}$, and $E=0.1$ V/\AA. (c,d) The corresponding $f$-wave spin-resolved isoenergy surfaces. (e,f) Electronic band structures of CoS and CoSe under EPL, with $\hbar\omega=10$ eV, $\tilde{A}_x=0.3$ \AA$^{-1}$, $\tilde{A}_y=0.2$ \AA$^{-1}$, and $E=0.1$ V/\AA; (g,h) the corresponding $f$-wave spin-resolved isoenergy surfaces. Red and blue denote spin-up and spin-down states, respectively. The isoenergy surfaces are plotted at selected energies of $-0.92$, $-1.03$, $-0.85$, and $-0.98$ eV for (c), (d), (g), and (h), respectively.}
	\label{fig6}
\end{figure*}
Beyond the light-induced odd-parity magnetism discussed above, we find that the combined application of light and an external electric field provides an additional route to realize mixed-parity magnetism in collinear antiferromagnets. Here, we use the term mixed-parity magnetism to describe a spin splitting that cannot be classified as purely even- or odd-parity~\cite{zhuang2026mixed}, i.e.,
\begin{equation}
	\varepsilon(s,\mathbf{k}) \neq \varepsilon(s,-\mathbf{k}),
	\qquad
	\varepsilon(s,\mathbf{k}) \neq \varepsilon(-s,-\mathbf{k}).
\end{equation}
As established previously, CPL and EPL induce odd-parity spin splittings with $f$-wave and $p$-wave characters in monolayer antiferromagnetic CoS and CoSe, respectively. We now consider superimposing these light fields with an out-of-plane electric field ($E=0.1$ V/\AA). In monolayer CoS and CoSe, the electric field drives the systems into a fFIM state with an approximately isotropic spin splitting (see the
SM~\cite{SM}). Simultaneous application of the electric field and polarized light yields a superposition of this isotropic fFIM background and the light-induced anisotropic component ($f$-wave for CPL, $p$-wave for EPL). As shown in Fig.~\ref{fig6}, this interplay substantially reconstructs the momentum-space textures, distorting the characteristic symmetries of the pure $f$-wave [Figs.~\ref{fig6}(a)--(d)] and $p$-wave [Figs.~\ref{fig6}(e)--(h)] states. These results establish that combining electric and light fields offers a versatile mechanism to tune spin-splitting textures: the electric field controls the isotropic background, while light determines the anisotropic angular dependence. This approach bridges pure odd-parity and mixed-parity states within a single platform, providing a robust pathway for the electrical and optical manipulation of unconventional compensated magnets. We further investigate the electronic structures of monolayer CoS and CoSe under the combined application of light irradiation, an external electric field, and uniaxial strain (see the SM~\cite{SM}). The results demonstrate that the induced spin splitting retains its mixed-parity character even when all three external perturbations are applied simultaneously.

\section{Light-induced anomalous Hall effect}
\begin{figure}[htb]
	\includegraphics[width=8.6cm]{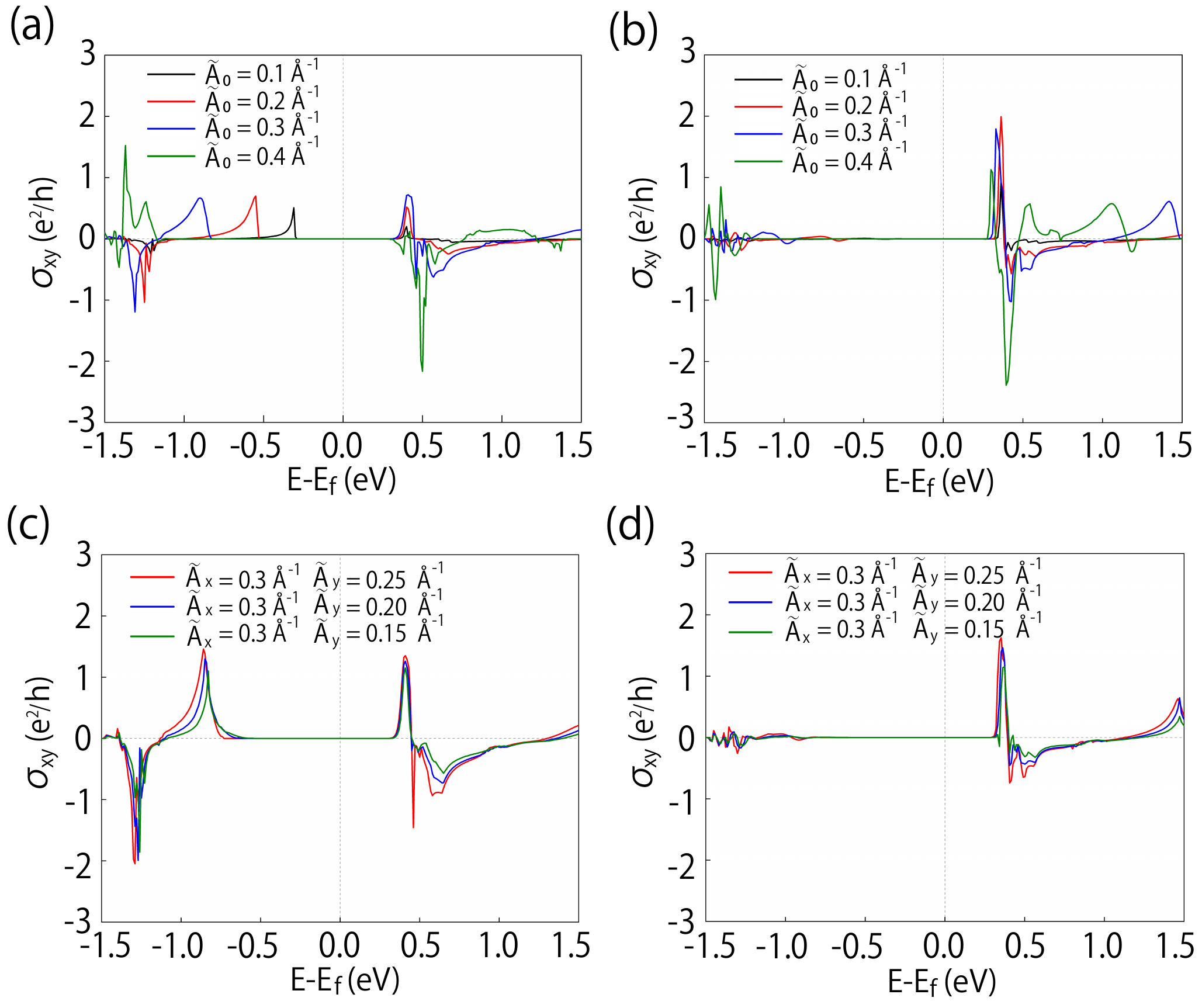}
	\caption{CPL-induced anomalous Hall conductivity of (a) CoS and (b) CoSe, and EPL-induced anomalous Hall conductivity of (c) CoS and (d) CoSe.}
	\label{fig7}
\end{figure}

In a collinear antiferromagnet with $\mathcal{PT}$ symmetry, the Berry curvature is constrained to cancel upon integration over the Brillouin zone, resulting in a vanishing anomalous Hall conductivity (AHC). When an external light field breaks the $\mathcal{PT}$ symmetry, this cancellation can be lifted, allowing a finite Brillouin-zone integral of the Berry curvature and, consequently, a nonzero AHC even in the absence of net magnetization. Within the Floquet formalism and first-principles calculations, the light-induced AHC in two-dimensional systems can be obtained within linear-response theory~\cite{jungwirth2002anomalous,yao2004first,oka2009floquet} as
\begin{equation}
	\sigma_{xy}
	=
	-\frac{e^2}{\hbar}
	\int_{\mathrm{BZ}}
	\frac{d^2k}{(2\pi)^2}
	\sum_n
	f\left(E_{n\mathbf{k}}\right)
	\Omega_{n,z}(\mathbf{k}),
	\tag{1}
\end{equation}
where $f\left(E_{n\mathbf{k}}\right)$ is the Fermi--Dirac distribution function for the $n$th band with energy $E_{n\mathbf{k}}$, and $\Omega_{n,z}(\mathbf{k})$ denotes the $z$ component of the Berry curvature of the $n$th band.
Figures~\ref{fig7}(a) and \ref{fig7}(b) show the anomalous Hall conductivity $\sigma_{xy}$ of monolayer CoS and CoSe, respectively, under CPL with different vector-potential amplitudes $\tilde{A}_{0}$. The photon energy is fixed at $\hbar\omega = 10~\text{eV}$, while $\tilde{A}_{0}$ varies from $0.1$ to $0.4~\text{\AA}^{-1}$. For both materials, the magnitude of $\sigma_{xy}$ increases with increasing light intensity. At $\tilde{A}_{0} = 0.4~\text{\AA}^{-1}$, the magnitude of $\sigma_{xy}$ reaches approximately $2e^{2}/h$. This sizable AHC demonstrates that light-induced odd-parity spin splitting can give rise to pronounced anomalous Hall responses in fully compensated collinear antiferromagnets.
To further elucidate the role of light polarization, Figures~\ref{fig7}(c) and \ref{fig7}(d) show the evolution of the AHC under EPL, with the vector-potential amplitude fixed at $\tilde{A}_{x}=0.3~\text{\AA}^{-1}$ and $\tilde{A}_{y}$ varied among $0.15$, $0.20$, and $0.25~\text{\AA}^{-1}$. The different curves correspond to distinct polarization states, characterized by the amplitude ratio $\tilde{A}_{y}/\tilde{A}_{x}$. It is found that the peak magnitude of the AHC increases with increasing $\tilde{A}_{y}/\tilde{A}_{x}$, indicating that the Hall response is enhanced as the polarization state evolves toward the circular-polarization limit.

\section{CONCLUSION}
In conclusion, we have demonstrated that light and external fields provide versatile routes to engineering odd- and mixed-parity spin splitting in fully compensated collinear antiferromagnets. Using monolayer CoS and CoSe as representative systems, we show that CPL induces odd-parity $f$-wave spin splitting, while EPL or uniaxial strain combined with CPL can realize the $p$-wave state. Moreover, the application of an electric field together with CPL or EPL converts the spin splitting into mixed-parity forms, offering additional flexibility for controlling the momentum-space spin texture. The light-induced breaking of $\mathcal{PT}$ symmetry, together with SOC, further generates a finite anomalous Hall conductivity without net magnetization. These results establish a direct connection between external-field-induced symmetry breaking, parity-dependent spin splitting, and anomalous transverse transport. More broadly, our findings provide a practical route for dynamically controlling momentum-space spin textures and anomalous Hall responses in compensated antiferromagnets, and may open new opportunities for light-controlled magnetic and spintronic applications.

\bigskip
\begin{acknowledgements}
	The authors thank Pu Zhang for valuable discussions. This work was supported by the National Natural Science Foundation of China (Grants No. 12204378, No. 12434006, and No. 12247103), the Major Basic Research Program of Natural Science of Shaanxi Province (Grant No. 2021JCW-19), and the Natural Science Basic Research Program of Shaanxi Province (Grants No. 2025JC-QYCX-007 and No. 2026JC-YXQN-026).	
\end{acknowledgements}


%

\end{document}